# Simulations of molecular logic gates

Kamil Walczak [1]

Institute of Physics, Adam Mickiewicz University
Umultowska 85, 61-614 Poznań, Poland

Here we propose four-terminal molecular devices as functional logic gates (AND, NOR and XOR, respectively). Such devices are composed of organic molecule connected to gold electrodes and located in between gate terminals. Specifically, the operation principle of molecular logic gates is discussed in detail. The input signals of the gate voltages can modify the output signal of the current carried through the device (for concrete value of applied bias) to simulate classical (Boolean) logical operations. Calculational scheme for current-voltage characteristics is based on the Landauer transport theory, where molecule is described with the help of tight-binding model, gold electrodes are treated by the use of effective Newns-Anderson theory and gate terminals are modeled as capacitor plates.

## I. Introduction

*Motivation*. Current computers are based on semiconductor logic gates which perform binary arithmetic and logical operations. However, trend of device miniaturization will reach its molecular-scale ultimate in the near future. Therefore the design and construction of molecular systems capable of performing complex logic functions is of great scientific interest now [1-7]. Reduction of size is not the only advantage offered by molecular electronics, but also the possibility of using quantum mechanical effects (such as: tunneling, quantum interference, quantization of molecular energy levels and discreteness of electron charge and spin) that would control the operation of nanodevices is very promising [8]. Several different methods have been proposed as possible realization of molecular logic devices.

*Preliminary proposals*. Molecular machines can perform simple logic operations, where fluorescence can be particularly useful signal to monitor such operations [3]. Another concept is associated with the light-driven molecular logical devices, in which the system is addressed and read by laser pulses [4]. Instead of analyzing interaction with light, there is still possibility to use electrical contacts. In the linear response regime of the current-voltage (I-V) behavior we can try to construct diode-based molecular electronic circuits [5]. However, novel devices may also take advantage of the current nonlinearity inherent to small molecules, allowing new architectures for computation that have no meaningful analogs in a conventional solid-state electronics [6,7].

*Purpose of this work*. Much effort has been done to understand transport phenomena into two-terminal devices, which may be suitable for simple memories [9] but not for more advanced circuits. However, we also should not forget about multi-terminal devices as potentially important for future applications. Recent theoretical reports were devoted to the question of transistor-like behavior relating to various kinds of three-terminal molecular devices [10-17]. In this paper we present an analytic approach to electron transport through four-terminal device, based on organic molecule connected to two gold electrodes and located in between plates of capacitor. Practical realization of the proposed device seems to be feasible by the use of presently available techniques.

*Conceptual solution*. Active molecule can be joined to the gold surfaces thanks to a mechanically controllable break (MCB) junction method [18,19] and gate terminals could be attached using the scanning tunneling microscope (STM) [20] by positioning a metallic tip close to the bridging molecule. So chemical bonds are realized only between sulfur atoms and gold electrodes [21], while the coupling with STM tips is assumed to be purely capacitive (there is no gate current in the system). In practice, the gate terminals can be fabricated in arbitrary form and size, that's why all the results in this paper are discussed in terms of gate voltage (however, it is also dictated for the sake of simplicity).

## II. Calculational scheme

*Four-terminal device*. Proposed device is composed of organic molecule connected to two gold electrodes and located in between plates of capacitor (gate terminals). These additional electrodes are used to control current flowing through the device (from the source to drain) under the influence of fixed bias voltage. Organic molecules are attached to gold electrodes by thiol end groups (–SH) that adsorb easily to the gold substrates (hydrogen atom of the thiol termination is desorbed, while sulfur atom bonds with three gold atoms arranged in an equilateral triangle [21]). In order to simplify our calculations we apply one-orbital approximation to the transport problem of molecular junctions. However, parameters of this simple scheme have been calibrated by matching with *ab initio* models. Therefore we believe that the essential qualitative physics and chemistry of the junction is captured.

*Description of molecular system*. Because only delocalized π-electrons are responsible for conduction through organic molecules, we decided to describe molecular bridges using simple version of tight-binding approximation (one π orbital per each carbon and sulfur atom). The carbon-carbon interaction energy is taken to be –2.5 eV and carbon site energy is chosen as –6.6 eV, which are standard parameters widely used to describe carbon conjugated structures [22]. The sulfur-carbon coupling of –1.5 eV is empirically determined to obtain a good fit to the *ab initio* energy levels [16], while sulfur site energy is equal to –6.6 eV because of its similarity in electronegativity with carbon atom.

*Contact with the electrodes*. The gold electrodes are analyzed through the use of one-orbital approximation, taking into account one s-type orbital per each gold atom. All the parameters of the model ($\beta = 0.4$ eV, $\gamma = 16.0$ eV) are taken after Damle *et al*. [16], because of some essential arguments included in their original work. Both of the electrodes are defined with the help of constant density of states (DOS) within energy bandwidth [23] and to estimate the order of magnitude of chemisorption coupling ($\Delta$) we use the simplified formula [24]: $\Delta = \beta^2 / \gamma = 0.01$ eV. However, our essential conclusions can be generalized beyond this approximation, and all the obtained results do not change qualitatively for a range of values of chosen parameters.

*Fermi energy*. Location of the Fermi energy relative to the electronic structure of the molecule is one of the most important factors in determining I-V characteristics of molecular conductors. When the molecule is coupled to the electrodes, there is some charge transfer even in the absence of the applied bias (molecule does not remain exactly neutral). In this way, the whole system (electrode-molecule-electrode) attains equilibrium with one Fermi level, which is located by requirement that the number of states below the Fermi energy must be equal to the number of electrons in the molecule [25]. Precise position of the Fermi level strongly depends on the contact model [23,25-27] and in our simplified approach it is treated as a parameter equal to – 5.1 eV to be in a good agreement with *ab initio* calculations [27].

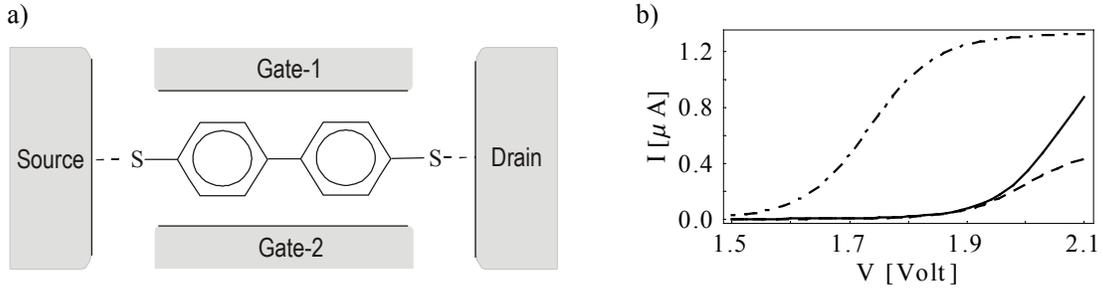

Figure 1. Schematic representation of molecule-based AND logic gate (a) and its I-V characteristics (b) for different gate voltages ($V_{G1} - V_{G2}$): (0-0) – dashed curve, (0-1, 1-0) – solid curve, (1-1) – dashed-dotted curve.

Table 1. Calculated current for a fixed bias ($V = 1.8$ Volt) given for several gate voltages in the case of molecular AND logic gate.

| $V_{G1}$ [Volt] | $V_{G2}$ [Volt] | $I$ [$10^{-6}$ A] |
|---|---|---|
| 0 | 0 | 0.02 |
| 1 | 0 | 0.02 |
| 0 | 1 | 0.02 |
| 1 | 1 | 1.01 |

*Gate terminals*. Gate electrodes are assumed to be ideally isolated from the molecular bridge (we can neglect the gate current only in a purely capacitive contacts) acting as two parallel plates of a capacitor. As a starting point of our simplified model we assume a flat potential profile in the region occupied by the molecule (which may not be true at high bias). By applying one or two control voltages to the molecule, one can modify the source–drain current flowing through the device. In our crude treatment, the gate voltages each operate on the nearest sites only, but in more sophisticated (self-consistent) models this effect should be extended over the other sites. However, this fact does not change our results qualitatively. In practice, one approach for including a controllable voltage would be to change the distance between gate terminal (STM tip) and bridging molecule.

*Transport theory*. To determine the current as a function of finite bias voltage applied between the source and drain electrodes (I-V characteristic) at finite temperature ($T = 300$ K), we use integration procedure of the transmission in the standard Landauer formulation [28]. Furthermore, transmission probability through a electrode-molecule-electrode system for an electron at given energy is calculated through the use of the non-equilibrium Green's function (NEGF) formalism [28]. In this way, all the simulations are based on the tight-binding version of molecular description (HMO level), effective Newns-Anderson treatment of molecule-to-electrodes coupling [29] and capacitor model of the gate terminals [30].

**III. Operation principle of molecular logic gates**

*Classical logic devices*. Generally, logic gates are switches whose output state (0 or 1) depends on the input conditions (also 0 or 1). In this section we show some proposals of complete logic devices (AND, NOR and XOR gates) based on individual molecules. In the near future it should be possible to fabricate not only single-gated molecular structures (FET-like devices) but also double-gated ones. Now the input signals are gate voltages ($V_{G1} - V_{G2}$), while the output is current flowing through the device ($I$) under the influence of the source-drain bias ($V$) – relatively small or large values of the current correspond to the output signals 0 or 1, respectively. The simplest logic devices are YES and NOT single-input gates. A YES gate passes the input bits to the output without changes, while a NOT inverts any input data.

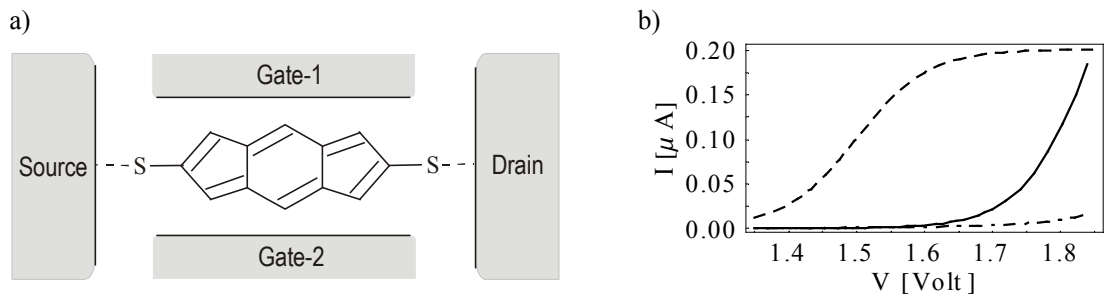

Figure 2. Schematic representation of molecule-based NOR logic gate (a) and its I-V characteristics (b) for different gate voltages ($V_{G1} - V_{G2}$): (0-0) – dashed curve, (0-1, 1-0) – solid curve, (1-1) – dashed-dotted curve.

Table 2. Calculated current for a fixed bias ($V = 1.6$ Volt) given for several gate voltages in the case of molecular NOR logic gate.

| $V_{G1}$ [Volt] | $V_{G2}$ [Volt] | $I$ [$10^{-7}$ A] |
|---|---|---|
| 0 | 0 | 1.74 |
| 1 | 0 | 0.04 |
| 0 | 1 | 0.04 |
| 1 | 1 | 0.02 |

*Single-input logic gates*. Simple YES gate can be simulated even by two-terminal molecular device, without using additional gate electrodes. When the applied bias is small (input 0), the current carried through the device is also small (output 0). If the voltage is high enough to open channels for electron transfer (input 1), then the current value is relatively high (output 1). One channel is opened when chemical potential of the electrode crosses one of the molecular energy levels [31]. As it is shown in table 2, NOR gate can be adopted readily to produce NOT gate by applying only one additional terminal (the voltage applied to the gate electrode can determine whether the molecular transistor does or does not allow a current to flow.). However, the NOR logic device is better than just a simple NOT gate, because the NOR makes it possible to construct some higher functions in a particularly simple manner.

*Double-input logic gates*. Figure 1b shows I-V characteristics of the system presented in figure 1a for several combinations of gate voltages. It is easy to see from this picture that for a range of source-drain bias 1.7-2.0 Volts we can observe significant current only in the case of (1-1) input signal. Such behavior is appropriate for AND logic gates (see also table 1). Similarly, in figure 2b we plot I-V dependences of the device configuration shown in figure 2a for several combinations of gate voltages. But this time we expect large values of the current only in the absence of gate voltages (input signal 0-0) for a range of source-drain bias 1.5-1.7 Volts. Indeed, that system shows the input/output relationships indicated by the truth table 2 of the NOR logic gate. Figure 3b presents I-V functions of the system schematically depicted in figure 3a (also for several combinations of gate voltages). Now the situation is essentially different. High values of the current are expected only when one gate is active for a range of source-drain bias 0.2-0.4 Volts. According to our predictions, such device should behave in accordance with the truth table 3 for a XOR logic gate. The on/off current ratios of all those logic gates are of order of magnitude of 30 to 100. It seems to be sufficient for experimental results to distinguish both options (low current – 0, high current – 1).

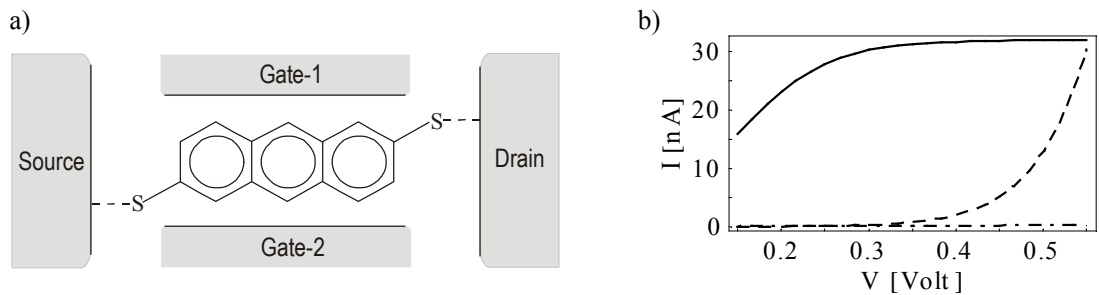

Figure 3. Schematic representation of molecule-based XOR logic gate (a) and its I-V characteristics (b) for different gate voltages ($V_{G1} - V_{G2}$): (0-0) – dashed curve, (0-1, 1-0) – solid curve, (1-1) – dashed-dotted curve.

Table 3. Calculated current for a fixed bias ($V = 0.3$ Volt) given for several gate voltages in the case of molecular XOR logic gate.

| $V_{G1}$ [Volt] | $V_{G2}$ [Volt] | $I$ [$10^{-8}$ A] |
|---|---|---|
| 0 | 0 | 0.03 |
| 1 | 0 | 3.02 |
| 0 | 1 | 3.02 |
| 1 | 1 | 0.02 |

## IV. Further perspectives

*Summary*. This work should be treated as the first step in understanding the operation of molecular logic gates, where the input signals are the gate voltages and output is the current flowing through the device for concrete value of applied bias. However, due to some limitations involved in our calculational scheme, obtained results should be considered as qualitative only. Further studies (more detailed self-consistent analysis) are still needed to confirm our predictions and to indicate all the factors that have significant influence on transport characteristics of four-terminal devices.

*Future challenges*. In our parametric approach Fermi energy was chosen arbitrarily and the study of its positioning will be performed in the supramolecular approach [23,25]. Moreover, electrostatic potential profile along the molecular bridge was postulated. However, because of the fact that electrostatics plays the key role in operation of FET-like devices, voltage drop inside the molecule will be calculated in the self-consistent manner [16]. In the frames of this work we also neglected the possible influence of inelastic effects and electron-electron interactions on transport phenomena in molecular devices.

*Conformational effects*. Another drastic simplification stems from the assumption that molecule is treated as a rigid structure, which does not deform in an external electric fields during the operation of the device (what can be justified only in the low-voltage regime). However, real molecules are capable of deforming in magnificent external fields [32] and it may be possible to take advantage of such conformational effects to design logic devices with superior characteristics [33].

*Open question*. How to connect all those molecular logic gates together to fabricate the integrated circuit (IC) ?